\documentclass[11pt]{article}
\usepackage{times}
\usepackage{mathfont}
\usepackage{cite}
\usepackage{url}

\long\def\comment#1{}

\DeclareSymbolFont{AMSb}{U}{msb}{m}{n}
\DeclareSymbolFontAlphabet{\Bbb}{AMSb}
\def\R{\ensuremath{\Bbb R}}
\let\Real\R

\def\F{\mathcal{F}}

\newcounter{bibnos}
\def\thebibliography#1{\list
  {[\arabic{enumi}]}{\settowidth\labelwidth{[100]}\leftmargin\labelwidth
  \advance\leftmargin\labelsep
  \usecounter{enumi}}%
  \setcounter{enumi}{\thebibnos}%
  \def\newblock{\hskip .11em plus .33em minus .07em}%
  \sloppy\clubpenalty4000\widowpenalty4000\small
  \baselineskip 10pt\itemsep-2pt%
  \sfcode`\.=1000\relax}

\begin{document}

\title{Emerging Challenges in Computational Topology}

\author{\parbox{5.25in}{Marshall Bern, David Eppstein,
Pankaj K. Agarwal, Nina Amenta, Paul Chew,
Tamal Dey, David P. Dobkin, Herbert Edelsbrunner,
Cindy Grimm, Leonidas J. Guibas, John Harer, Joel Hass, Andrew Hicks,
Carroll K. Johnson, Gilad Lerman, David Letscher,
Paul Plassmann, Eric Sedgwick, Jack Snoeyink,
Jeff Weeks, Chee Yap, and Denis Zorin
\thanks{Author affiliations:
Marshall Bern, Xerox Palo Alto Research Ctr., bern@parc.xerox.com.
David Eppstein, Univ.\ of California, Irvine, Dept.\ of Information \&
Computer Science, eppstein@ics.uci.edu.
Pankaj K. Agarwal, Duke Univ., Dept.\ of Computer Science,
pankaj@cs.duke.edu.
Nina Amenta, Univ.\ of Texas, Austin, Dept.\ of Computer Sciences,
amenta@cs.utexas.edu.
Paul Chew, Cornell Univ., Dept.\ of Computer Science, chew@cs.cornell.edu.
Tamal Dey, Ohio State Univ., Dept.\ of Computer and Information Science,
tamaldey@cis.ohio-state.edu.
David P. Dobkin, Princeton Univ., Dept.\ of Computer Science,
dpd@cs.princeton.edu.
Herbert Edelsbrunner, Duke Univ., Dept.\ of Computer Science,
edels@cs.duke.edu.
Cindy Grimm, Brown Univ., Dept.\ of Computer Science, cmg@cs.brown.edu.
Leonidas J. Guibas, Stanford Univ., Dept.\ of Computer Science,
guibas@cs.stanford.edu.
John Harer, Duke Univ., Dept.\ of Mathematics, harer@math.duke.edu.
Joel Hass, Univ.\ of California, Davis, Dept.\ of Mathematics,
hass@math.ucdavis.edu.
Andrew Hicks, Univ.\ of Pennsylvania,
Dept.\ of Computer \& Information Science, rah@grip.cis.upenn.edu.
Carroll K. Johnson, Oak Ridge National Lab., ckj@ornl.gov.
Gilad Lerman, Yale Univ., Dept.\ of Mathematics, gilad.lerman@yale.edu.
David Letscher, Univ.\ of California, San Diego, Dept.\ of
Mathematics, letscher@math.ucsd.edu.
Paul Plassmann, Pennsylvania State
Univ., Dept.\ of Computer Science \& Engineering, plassman@cse.psu.edu.
Eric Sedgwick, Oklahoma State Univ., Dept.\ of Mathematics,
sedgwic@math.okstate.edu.
Jack Snoeyink, Univ.\ of North Carolina,
Chapel Hill, Dept.\ of Computer Science, snoeyink@cs.unc.edu.
Jeff Weeks, weeks@northnet.org.
Chee Yap, New York Univ., Dept.\ of Computer Science, yap@cs.nyu.edu.
Denis Zorin, New York Univ., Dept.\ of Computer Science,
dzorin@mrl.nyu.edu.}}}

\date{}

\maketitle

\vspace{.5in}

\begin{abstract}
{\normalsize
Here we present the results of the NSF-funded Workshop on Computational
Topology, which met on June 11 and 12 in Miami Beach, Florida.
This report identifies important problems involving both 
computation and topology. 
}
\end{abstract}

\newpage

\section{Introduction and Background}

\hrulefill\par\bigskip\noindent

Over the past 15 years,
computational geometry has become a very
productive area, with applications in fields such as graphics,
robotics, and computer-aided design.  Computational geometry, however,
primarily focuses on discrete problems involving point sets,
polygons, and polyhedra, and uses combinatorial techniques
to solve these problems.
It is now time for computational geometry to broaden its scope in order
to meet the challenges set forth in the President's
Information Technology Advisory Committee (PITAC) Report
\cite{PITAC}
and the Information Technology for the Twenty-First
Century (IT$^2$) Initiative \cite{IT2},
specifically the need for accelerated progress in
information visualization, advanced scientific and engineering
computation, and computational algorithms and methods.

There is a need to extend computational geometry---with
its emphasis on provable correctness, efficiency, and robustness---to
continuous domains, curved surfaces, and higher dimensions.  
Such an extension brings computational geometry into contact
with classical topology, just as earlier research 
led to inextricable connections with combinatorial
geometry---to the great benefit of both fields.

We intend the name
{\it computational topology\/}
to encompass both algorithmic questions
in topology (for example, recognizing knots)
and topological questions in algorithms
(for example, whether a discrete construction preserves
the topology of the underlying continuous domain). 

Research into computational topology has started already
\cite{Veg-HDCG-97}, and is at present being undertaken separately by
topology, computational geometry, and computer graphics communities,
among others.  Each of these fields has developed its own favored
approaches to shape representation, manipulation, and analysis. 
Algorithms are often specific to certain data representations,
and the underlying questions common to all approaches
have not been given adequate attention.

The Workshop on Computational Topology, June 11--12, 1999,
in Miami Beach, Florida,
brought together researchers involved in aspects of
computational topology. 
The purposes of this interdisciplinary workshop
were to set goals for computational topology,
identify important problems and areas,
and describe key techniques common to many areas.

\newpage

\section{Goals}

\hrulefill\par\bigskip\noindent

Geometric computing is a fundamental element in several
of the areas highlighted in the IT$^2$ initiative:
information visualization \cite[section 2, ``Fundamental Information
Technology Research'', under human-computer
interaction and information management]{IT2}, advanced science and
engineering computation \cite[section 3, ``Advanced Computing for
Science, Engineering, and the Nation'']{IT2}, and computational and
algorithmic methods \cite[section 3, under computer science and
enabling technology]{IT2}. Scientific and engineering computing often
simulates physical objects and their interactions, on scales that vary
from the atomic to the astronomical. Modeling the shapes of these
objects, and the space surrounding them, is a difficult part of these
computations. Information visualization also involves shapes and motions,
as well as sophisticated graphics rendering techniques.
Each of these two areas, as well as many others, would benefit
from advances in generic computational and algorithmic methods.

Some of the most difficult and least understood issues in
geometric computing involve topology.
Up until now, work on topological issues
has been scattered among a number of fields, and its level
of mathematical sophistication has been rather uneven.
This report argues that a conscious focus on computational topology
will accelerate progress in geometric computing.

Topology separates global shape properties from local geometric attributes,
and provides a precise language for discussing these properties.
Such a language is essential for composing software programs, such
as connecting a mesh generator to a computational fluid dynamics
simulation.
Mathematical abstraction can also unify similar concepts
from different fields. For example,
basic questions of robot reachability or molecular docking
become similar topological questions in the appropriate configuration or
conformation spaces.
Finally, by separating shape manipulation from app\-li\-ca\-tion-specific
operations, we expect to improve reliability of geometric computing
in many domains,
just as other large software systems (for example,
operating systems and internet routing)
have gained reliability through layered design.

\newpage

\section{Areas and Problems}\label{sec-areas}

\hrulefill\par\bigskip\noindent

We have identified five main areas in which computational
topology can lead to advances in simulation and visualization.
\begin{itemize}
\item
{\bf Shape acquisition.}
The entry of the
shapes of physical objects into
the computer is becoming increasingly automated.
Part of this process is developing algorithms that turn a
set of measurements or readings into a topologically valid
shape representation.

\item
{\bf Shape representation.}
Many different computer representations of shape are in use.
Describing the relationships between them, converting from one
to the other, and developing new representations,
all require topological ideas and methods.

\item
{\bf Physical simulation.}
For scientific and engineering computations, shape representations
are typically meshed into small pieces.
Many of the issues that have arisen in mesh generation
are topological.

\item
{\bf Configuration spaces.}
Configuration or conformation spaces represent the possible
motions of objects moving among
obstacles, mechanical devices, or molecules.
These spaces are usually high-dimensional and non-Euclidean,
and hence raise some rather deep topological questions.

\item
{\bf Topological computation.}
Some recent advances in topology itself involve algorithms and
computation. Better software for geometric computing will help advance
this approach to topology,
while new techniques and representations  developed
for topological problems will contribute to the advancement of
geometric computing.

\end{itemize}

In each area we have selected  a few problems for more detailed
discussion.

\subsection{Shape Acquisition}

Computer representations of shapes
can either be designed by a person using CAD tools or acquired from an 
existing physical object.
The latter approach offers advantages of speed and faithfulness to an
original, which is of course crucial in applications such as medical
imaging. Automatic acquisition of shapes poses a wealth
of geometric and topological challenges.

\paragraph{Shape Reconstruction from Scattered Points.}

Modern laser scanners can measure a large number of points on
the surface of a physical object in a matter of seconds.  The
most basic computational problem is the reconstruction
of the ``most reasonable'' geometric shape that generated the point sample.
We find algorithmic solutions to this problem in both the
computational geometry and the computer graphics literature.
Consistent with the dominant cultures in these two
areas, solutions suggested in computational geometry are
discrete in nature, while solutions in computer graphics are
based on numerical ideas.

The early work of Hoppe et al.~\cite{HopDeRMcD-CG-92} in computer graphics
drew wide attention to the reconstruction problem.  The authors
give an algorithm that works for point data sampled everywhere
densely on the surface of the object.  A basic step in the
reconstruction estimates the surface normal at a point using a
best-fit plane determined by near neighbors.  This idea is
inherently differential and limits the algorithm to shapes and
data for which locally linear approximations provide useful
information.  Raindrop Geomagic, Inc.  takes
a more global approach in its software Wrap,
which reconstructs a shape using the 3-dimensional Delaunay
triangulation of the sampled points~\cite{Wrap}.
Amenta and Bern describe another algorithm using the
Delaunay triangulation together with differential
ideas~\cite{AmeBer-SCG-98}, and go on to prove that their algorithm
gives a geometrically close and topologically correct output,
under certain assumptions about the point sample.
This result rationalizes the reconstruction process and focuses 
attention on the more difficult cases in which the assumptions are violated,
for example, surfaces with creases or boundaries, and sample points with noise.

\paragraph{Manifold and Space Learning. }

Mathematically, it makes perfect sense to generalize the
reconstruction problem from ${\R}^3$ to higher-dimensional
Euclidean space.  Perhaps somewhat surprisingly, this
generalization makes sense also from the viewpoint of applications,
including speech recognition, weather forecasting, and economic prediction.
Many natural phenomena can be sampled by individual
measurements, where each measurement can be interpreted as a point
in ${\R}^d$, for some fixed dimension $d$. 

The reconstruction problem in ${\R}^d$ is more difficult
than in ${\R}^3$ not only because $d$ usually exceeds 3, but also
because we have no a priori knowledge about the intrinsic
dimension of the shape that we wish to reconstruct.  It might
have mixed or fractal, or even altogether ambiguous, dimension.
Since the input data is a discrete set of points,
which by definition has dimension zero, the question itself is
highly ambiguous and the answer depends on the scale at which we
view the data.  The idea of scale dependent variation as applied
in the definition of fractal or Hausdorff dimension~\cite{Mat-95} thus
suggests itself.  It appears in the work of Jones~\cite{Jon-IM-90}, where
local dimension is estimated through the variation of linear
best-fits in a hierarchy of nested neighborhoods.  It is also
manifest in the work of Edelsbrunner and M\"{u}cke~\cite{EdeMue-TOG-94},
who define alpha shapes as a family of reconstructed shapes
parametrized by scale.  One of the challenging problems in this
context is the study of the interaction between noise and scale.

\paragraph{Reconstruction from Slices. }

In many applications the input data includes additional information
that can help in reconstructing the shape.  Examples
are estimates of surface normals provided by the scanner or
information encoded in the sampling sequence.  A classic
version of the reconstruction problem in the latter category
presents the data in {\it slices\/}, each slice consisting of one
or more polygons given by a cyclic sequence of vertices.
Usually, the slicing planes are parallel, in which case the
reconstruction reduces to connecting each pair of contiguous slices.

A problematic aspect of this approach to shape reconstruction is
the nonsymmetric treatment of coordinate directions.
In other applications, however, nonsymmetric treatment
seems warranted or even necessary.  For example, the animation
of a moving planar shape can be viewed as sweeping a surface
in ${\R}^2$ times time.  
Branching occurs at critical points, which correspond to moments
in time where the shape changes its topology.  
The relevant mathematics here is 
Morse theory, which studies the combinatorial and
differential structure of critical points~\cite{Mil-63}.

A related problem is morphing: 
given two surfaces in ${\R}^3$, construct a continuous
deformation of one surface into the other.  The deformation may
be a homotopy or a cobordism.  Again we can view this as a
problem in Morse theory, only one dimension higher than before.
This view is adopted in~\cite{CheEdeFu-PCCGA-98}, where canonical
deformations are used in the construction of ``shape spaces''.  Such
spaces could be useful in building databases of shapes, such as
drug compounds, anatomical structures, and mechanical tools.

\paragraph{Crystal Structures from X-ray Data.}

The standard tool for
determining conformations of atoms and molecules in crystals
is X-ray crystallography. Missing phase information must be
inferred to convert observed Bragg diffraction intensity data
into a phased Fourier amplitude set.
This process continues to become more routine even
for macromolecules such as proteins and viruses. A Fourier transform of
the amplitude set then produces an electron density function over
$\R^3$. If the observed intensity set extends far enough, the peaks of
the the density function provide starting atomic coordinates suitable for
least squares refinement, however macromolecular density maps usually do
not have atomic resolution.

It is again convenient to describe the situation using the
language of Morse theory.  The density is viewed as
the height function of a 3-manifold in ${\R}^4$, with four types of
critical points:
``peaks'', ``passes'', ``pales'', and ``pits''.   The reconstruction of
the atomic or molecular configuration may be complicated by the presence
of excessive noise, thermal motion, positional and occupancy disorder, or
lack of atomic resolution.
Morse theory interpretations of crystallographic density
functions are being carried out for a variety of crystal
structures ranging from macromolecules with less than atomic
resolution \cite{ForChiGla-MC-97}, to ultra precise
small molecule structures for which quantum mechanical
perturbations such as lone pair density peaks in the middle of
covalent bonds are detectable \cite{Bad-94}.  When thermal
motion is the primary focus, neutron Bragg diffraction data
often are used, from which nuclear density rather than electon
density maps are produced and thus do not include quantum
perturbations.

Advances in computational topology can contribute to the above
and to other problems such as classification schemes for
crystal structures using Heegaard level surfaces between passes
and pales \cite{Joh-TMP-99}, and certain related minimal
surfaces \cite{LeoNes-IUCr-99}.
Delaunay-based reconstruction
might provide useful
tools to address the problem of ``topological noise''
such as spurious peaks and passes.
Morse theory can also play a role in efficient
algorithms for finding structures in electron density data
\cite{CarSnoAxe-TR-99}.

\subsection{Shape Representation}
 
Data structures for representing shapes have emerged independently in
many different fields. These representations include unstructured
collections of polygons (``polygon soup''), polyhedral models,
subdivision surfaces, spline surfaces, implicit surfaces, skin surfaces,
and alpha shapes.  Generally speaking, these methods are at best
adequate within their own fields,
and not well-suited for connecting across fields.
At least in the CAD area, 
there is growing awareness that future systems must be more
mathematically sophisticated than today's systems.
Rida Farouki writes, ``At the heart of this problem
lie some deep mathematical issues, concerned with the computation,
representation, and manipulation of complex geometries''
\cite{Far-SN-99}.
Shape representation appears to be an ideal area 
for collaboration between mathematicians and computer scientists.

\paragraph{Conversion Between Different Representations.}

A number of {\it ad hoc\/} methods exist for converting between 
different types of representations, usually to and from polyhedral
models.  These methods typically use geometric criteria to
evaluate the conversion, for example, guaranteeing that the original
surface is pointwise not more than a small tolerance distant from the polygonal mesh.
Geometry alone is insufficient, however, as it does not guarantee
topological properties such as ``watertightness''.
Including topological criteria in the evaluation will
lead to more correct conversion programs.

\paragraph{Topology Preserving Simplification.}
The process of replacing a polygonal surface with a simpler one,
while essential to many hierarchical representations, is notorious
for introducing topological errors which can be fatal for later
operations. A popular method, edge contraction
\cite{HopDeRDuc-SG-93}, can be applied to general simplicial complexes, but
is not in general guaranteed to produce a complex
homeomorphic to the original.
Dey et al.~\cite{DeyEdeGuh-TR-98}, however, have proved that the complex
after contraction is homeomorphic to the original if the neighborhood of
the contracted edge satisfies a {\em link condition}.
For 2- and 3-manifolds, the link condition defines the contractable edges.

Even if the output of a simplification process is homeomorphic to the
input, however, there is no guarantee that the output is
correctly embedded. Self-intersections are
often introduced, for instance, a problem sometimes know as ``bubbling''.
One step towards guaranteeing a correct embedding was
a paper by Varshney et al.~\cite{CohVarMan-SG-96}, in which a simplified
2-manifold is fitted into a shell around the original.
Much more work, both in providing mathematically verifiable guarantees
and in developing efficient algorithms, is required.

\paragraph{Smoothness and  Nonsmoothness.}

Smooth surface representations are commonly divided
into implicit and explicit representations.
Implicit surfaces can be defined by blending parametrized surfaces
such as splines, or as level sets of scalar functions.
The advantages of implicit surfaces include high degree of smoothness for
arbitrary topology, ease of raytracing, and ease of combining
several objects by blending. Disadvantages 
include difficulties with parameterization and conversion to polygonal
meshes. Moreover, implicit surfaces may have singularities, 
which can be difficult to detect and control.

The most common explicit representation---ubiquitous in
CAD---is that of nonuniform rational B-spline (NURBS) patches. 
A more systematic approach, which offers the advantage of
multiresolution control, involves subdivision surfaces~\cite{Zor-SG-99}.
Both of these methods, however, produce surfaces with defects, for
example, flat spots or areas of relatively low smoothness
near extraordinary points.
Whether or not a point is extraordinary depends on the local
topology within the representing mesh, and has nothing to do
with its geometric location.  The changed amount of smoothness
is thus an artifact of the representation, and should ideally
not exist.
An important challenge in smooth surfaces is to ensure integral measures of
visual smoothness (fairness). Variational surfaces aim to handle 
such measures directly.

In the other direction, there is also need for representations
that can handle singularities such as boundaries, creases, and corners.
With standard polyhedral models, there is no distinction
between the creases resulting from discretization and those that represent
true surface features.  
Spline patches give rise to creases of high algebraic degree that
cannot be manipulated directly, and 
implicit surfaces rarely allow any control of singularities.

\paragraph{Multiscale Representations.}
Multiscale representations, whether implicit or explicit, 
hold out the promise of efficiency even for very complex geometries. 
We identify the main challenge as
developing representations that allow controlled topology changes
between  levels, while supporting a variety of efficient
multiscale operations, such as animation, editing, and ``signal processing''.

Implicit multiscale representations (level sets) have been
used in volume rendering.  Volume data are themselves
represented on regular or adaptively refined (octree) grids; thus
it is natural to use classic  functional  multiscale
representations such as
wavelets~\cite{WesErt-CGF-97,Wes-SVV-94}, yet it is also
possible to construct unstructured mesh hierarchies on volume
data~\cite{WeiJoh-TR-95}.  While allowing topological changes at
different levels of the hierarchy,  implicit representations offer little
control of such changes.  On the other hand, at least in the case
of volume data represented
on regular grids, signal processing techniques can be used to
handle some of the topological problems~\cite{Wes-SVV-94}.

Some of the current explicit methods 
\cite{SchZarLor-CG-92,RosBor-GMCG-93,GarHec-SG-97} do
allow topological changes, but the control over such changes is relatively
limited and the hierarchies created by these methods are unsuitable
for many purposes, for example, it may be difficult or impossible 
to parameterize finer levels of hierarchies over the coarse levels.
The recent work of El-Sana and Varshney \cite{ElSVar-TVCG-98}
based on alpha shapes \cite{EdeMue-TOG-94}  aims to perform
topological simplification in a more controlled manner.

\paragraph{Qualitative Geometry and Multiscale Topology.}

For the final highlighted problem, we move from shape representation
to shape analysis.
Topological invariants (see Section~\ref{sec-topology}) such as Betti
numbers are insensitive to scale, and do not distinguish
 between tiny holes and large ones.
Moreover, features such as pockets, valleys, and ridges---which
are sometimes crucial in applications---are not
usually treated as topological features at all. 
Nevertheless, topological spaces naturally associated with a given surface 
can be used to capture scale-dependent and qualitative geometric features.

For example,
the lengths of shortest linking curves~\cite{DeyGuh-JACM-98},
closed curves through or around a hole, can be used
to distinguish small from large holes, and
the areas of {\it compressing disks\/}, which ``seal off'' a hole,
can be used to distinguish long narrow pipes from direct openings.
The topology of offset or ``neighborhood'' surfaces is an appropriate tool 
for classifying depressions in a surface: a sinkhole with
a small opening will seal off as the neighborhood grows,
whereas a shallow puddle will not.
Edelsbrunner has already used this idea to design an algorithm
to detect pockets in molecular surfaces~\cite{EdeFacLia-DAM-98}, but
further investigations are necessary to answer questions 
on the border of geometry and topology.

\subsection{Physical Simulation}

Scientific computing has traditionally been concerned with
numerical issues such as the convergence of discrete
approximations to partial differential equations (PDEs),
the stability of integration methods for time-dependent 
systems, and the computational efficiency of software
implementations of these numerical methods.
Of central importance has been ultimate use of these
techniques in the solution of complex problems in
science and engineering such as the modeling of combustion
systems, aerodynamics, structural mechanics, 
molecular dynamics, and problems from a large number of
other application areas.
However, as these applications have become more complex, the 
local convergence properties of numerical methods have not proven
to be sufficient to ensure either correctness or robustness.
There are a number of research 
areas where topological and differential methods could
be integrated with existing numerical techniques in scientific
computing to help resolve these difficulties.


\paragraph{Hexahedral Mesh Generation.}

For many scientific applications, the preferred discretization
is a {\em hexahedral mesh} partitioning the domain into cuboids. 
A common approach to hexahedral mesh generation involves 
extending a quadrilateral mesh on the domain surface
to a three-dimensional volume mesh of the entire 
domain~\cite{BerPla-TR-97}.
Even though several software implementations of this 
approach exist~\cite{TauMit-IMR-95}, it is not yet known whether 
this extension can always be done.
An obvious necessary condition for the existence of a hexahedral
mesh is that there be an even number of boundary quadrilaterals;
this is also sufficient to
guarantee the existence of a {\it topological\/} mesh,
meaning one in which hexahedral faces may be slightly 
nonplanar, for domains forming a simple topological
structure~\cite{Mit-STACS-96,Thu-93} or having a bipartite boundary
\cite{Epp-CGTA-99}.
However, it is not clear whether similarly simple conditions can
guarantee the existence of a polyhedral mesh
or whether additional algebraic conditions
on the surface must be imposed.

Another important issue in automatic
mesh generation is element quality; 
poorly shaped elements (flat or skinny, especially
skinny in the ``wrong direction'') 
are directly responsible for poorly
conditioned matrices~\cite{Fri-AIAA-72} and
hence slow and inaccurate numerical computations~\cite{BabRhe-SJNA-78}.
For triangular, tetrahedral, and quadrilateral meshes,
the solution to poor quality elements has been the introduction
of ``provably good'' meshing methods that guarantee to produce
a mesh with all elements having good quality according to various metrics
\cite{BerEpp-IMR-97,BerEppGil-JCSS-94}.
However for hexahedral meshes, little is known about 
quality metrics and even less is known about provably good meshing.
Recent work using the Jacobian matrix norm 
as a quality metric for hexahedral elements has shown
promise for finite-element calculations~\cite{Knu-TR-99-II}.

\paragraph{Anisotropic Mesh Generation.}
In many applications the underlying physics is not isotropic
and, as a result, standard mesh generation methods
and element quality metrics are not appropriate.
This is the case, for example, in modeling the fluid flow in 
a boundary layer, or in groundwater flow calculations where 
the porosity is highly nonisotropic because of geological
features such as faults and layering of strata.
For these problems element aspect ratios of 1000:1 are sometimes 
necesssary; however, the generation of these
meshes is often {\it ad hoc\/}.
For isotropic problems the shape optimization of elements
based on measures generated from local metrics computed
from the Hessian of element error functions has proved
useful \cite{Rip-SJNA-92} and optimal for the finite-element
approximation of given functions \cite{Sim-ANM-94}.
A promising area of future research is the extension of
these results to anisotropic problems.
For example, one would like to characterize the existence of
canonical triangulations (perhaps something
akin to Delaunay triangulation)
given the Riemannian metrics generated by the error function estimates.

\paragraph{Moving Meshes.}
In problems such as casting and molding, the domain 
changes with time, and it is convenient 
to adapt the existing mesh rather than recomputing an entirely new mesh.
Moving mesh problems also arise in Lagrangian discretization
strategies for time-dependent PDEs.
The challenge problem here is the
identification and correction  of topological changes
as the mesh changes over time.

\paragraph{Visualization.}
Large-scale simulations can generate terabytes of numerical data.
The analysis and interpretation of this voluminous data has become
an increasingly important research problem.
One promising approach extracts features 
such as vortex lines or sheets~\cite{SilZabFer-SVNP-93}.
The topology and qualitative geometry of these features
can be of great interest.
Examples include identification of voids and pockets
in  molecular surfaces~\cite{Bad-94,EdeFacLia-DAM-98}, and
simulation of high-temperature superconductors, 
in which magnetic field lines ``tangle''
with impurities in the material~\cite{GroKapLea-JCP-96,JonPla-IJSA-93}.

\subsection{Configuration Spaces}

The notion of \emph{configuration space} (also called
\emph{parametric space} or \emph{realization space}) is used
in numerous areas, including robotics, graphics, molecular
biology, computer vision, and databases for representing the
space of all possible states of a system characterized by many
degrees of freedom.
Instead of  defining  configuration spaces in general, we
will illustrate the concept by giving an example in robotics.

In \emph{robot motion planning},
the problem is to compute a collision-free motion
between two given placements---or {\it configurations\/}---of a given
robot among a set of obstacles. A configuration is typically described as
a list of real parameters, and the set of all possible configurations is
called the configuration space.
\emph{Free configuration space} $\F$ is
the subset of the configuration space at which the robot does not
intersect any obstacle. The robot can move from an initial
configuration to a final configuration  without intersecting any obstacle
if and only if these two configurations lie in the same connected
component of free configuration space.
Planning a collision-free motion
thus maps to planning the motion of a point in $\F$. In other
words, the motion-planning problems map to connectivity
questions, or related topological questions, in $\F$.
Many other problems can be couched in terms of configuration
spaces.  Important examples include {\it assembly planning\/} and
{\it molecular docking\/}~\cite{HalKavLat-HDCG-97,HalLatWil-97,Lat-91}.
The topology of configuration spaces is little understood,
except in very rudimentary cases, such as that of an
object under rigid motion.

\paragraph{Representation and Computation.}

Most interesting configuration spaces are {\it semialgebraic sets\/},
finite Boolean combinations of solution sets of
polynomial inequalities and equalities.
The question of representing
and computing a semi-algebraic set has received much
attention in the last two decades. Since the topology of
a semi-algebraic set can be quite intricate,
developing a suitable representation
is a challenging (and not fully solved) problem.
A common technique to represent a
semialgebraic set is to partition it into  semialgebraic
sets of constant description complexity,
each of which is homeomorphic to $\Real^j$ for some~$j$
\cite{Bri-DCG-93,Lie-CAD-91,SchSha-AAM-83}.
Some commonly used general decomposition schemes are Collin's
decomposition~\cite{ArnColMcC-SJC-84,Col-ATFL-75} and vertical
decomposition~\cite{ChaEdeGui-ICALP-89}. Because of
efficiency considerations, we want to minimize the number
of cells in the decomposition.
A major open question in this area is to compute a
decomposition of minimum  size.

In motion planning, we are interested in computing a single
connected component of $\F$. (It is not even obvious that a
connected component of a semialgebraic set is also
semi-algebraic; this was proved only recently~\cite{BasPolRoy-98}.)
What is the combinatorial or topological complexity of such a
component?
Recently, Basu proved tight bounds on the sum of Betti numbers
and used it to prove a sharp bound on the combinatorial
complexity of a single component~\cite{Bas-FOCS-98}.
However, no efficient algorithm is known for computing a single cell.
A related open problem is to develop an efficient
stratification scheme for a single component of a
semialgebraic set.

In some applications even more challenging problems arise.
If the obstacles are moving as well as the robot, then
we need to update a road-map or stratification  dynamically.
Also, flexible objects such as elastic bands, rope, or cloth
cannot be properly represented with a finite number of degrees of
freedom. How can we represent configuration spaces of such objects?
The key here may be to capture the notion that different
configurations have an energy associated with them, and that only
low-energy configurations are of interest. Are there good ways to
parametrize these low-energy configurations and to plan motions among
them?

\paragraph{Approximation.}

Computing exact high-dimensional configuration spaces
is impractical.  Thus it is reasonable to ask
for approximate representations.
Much of the difficulty in approximating a high-dimensional configuration
space is in understanding and simplifying the
topology of the space. Although several algorithms are known for
simplifying the geometry of a surface,
little is known about simplifying topology.

Recently, Monte Carlo algorithms have been developed for
representing a higher dimensional semialgebraic set by a
$1$-dimensional network~\cite{KavLatMot-STOC-95,KavSveLat-TRA-96}.
Intuitively, this network is an approximate
representation of the {\em road map}
(a network of $1$-dimensional curves that
captures the connectivity information of $\F$).
These methods sample points in $\F$ and connect them by
an edge  if they can be connected by a direct path
inside $\F$. So far very simple strategies have been developed
for choosing random points. These methods
work well when $\F$ is simple, but better sampling techniques
are needed to handle planning problems involving
narrow corridors or other difficult areas,
in such a way that the connectivity of the sampled configuration space
is preserved.

\paragraph{Decomposition.}

Dimension reduction is one approach to developing faster
algorithms for problems in high dimensions.
One possibility for motion planning is to search for solutions
in one or more projections of the configuration space and then lift
the solution back to the original space.
For example, suppose we want to plan a motion
for two disks in the plane amid obstacles.
The four-dimensional free space of this system can
be computed by decomposing the two-dimensional
free space of each disk into simple cells, and then lifting these cells
into $\Real^4$.
Proving that such a strategy succeeds requires
several sophisticated techniques from algebraic topology,
including  Mayer-Vietoris sequences \cite{AroBerSta-SCG-98,
ForWilYap-ICRA-86, HopWil-IJRR-86, HopWil-SJC-86}. A characterization of
the sitations in which the configuration space can be decomposed and
finding the ``optimal'' decomposition of the configuration space are two
interesting open problems in this area.

\subsection{Topological Computation}\label{sec-topology}

The study of algorithms for topological problems
has grown quite popular in recent years; it is one of
the few growing branches of topology.
In the last few years, there have been several workshops
and the founding of an on-line community---www.computop.org.
Much of the recent effort has focused on classifying the inherent
complexity of topological problems.
Typically, planar problems are easy (polynomially solvable),
problems in $\Real^3$ are hard (exponentially solvable and thought
to be NP-complete), and problems in $\Real^4$ and higher dimensions are
known to be undecidable.

\paragraph{Unknot Recognition.}

A knot is said to be unknotted if
it can be deformed to a (geometric) circle without passing through
itself.  In the early 1960s, Haken used a combinatorial representation
of surfaces, called {\it normal surfaces\/},
in an algorithm for deciding if a knot is unknotted~\cite{Hak-AM-61}.
A recent collaboration between mathematicians and computer scientists showed
that this algorithm will take at most exponential time in the
number of crossings of the knot~\cite{HasLagPip-JACM-99}.
It is still open, however, whether this problem is NP-complete
or can be solved in subexponential time.

\paragraph{Knot and Link Equivalence.}

Two knots are equivalent if one
can be deformed into the other without passing through itself.
Knot equivalence is known to be decidable~\cite{Hem-92},
but the algorithm is extremely complicated and the computational
complexity is as yet unknown.
An important related question asks whether two links (collections of 
intertwined knots) are equivalent.  No algorithm is yet known
for this problem.

\paragraph{Three-Sphere Recognition.}

The development of almost normal
surfaces, a generalization of normal surfaces, has led to the
Rubinstein-Thompson algorithm for deciding if a manifold is the
$3$-sphere \cite{Tho-MRL-94}.  Recent work of Casson shows that this
algorithm will take at most exponential time.

\paragraph{Shellings.}

A {\it shelling\/} of a cell complex is an 
ordering of the cells such that if cells are added one
by one in that order the topological type remains invariant.
While interesting for their own sake, 
shellings also provide a very useful calculational tool.
Hence it is an important algorithmic problem to determine if a cell complex
is shellable and, if not, modify it so that it is.
Current algorithms~\cite{Let-99} are not yet practical,
and improvements are needed. 

\paragraph{Hyperbolic Geometry.}

Three-dimensional manifolds with 
a hyperbolic structure have many useful properties,
allowing extremely efficient and powerful topological calculations. 
The computer software package SnapPea \cite{SnapPea},
written by Weeks, implements many of these calculations and has
proven an exteremely useful tool for low-dimensional topology.
There are still many open questions in algorithmic hyperbolic geometry,
for example, whether it is possible to decide if a
manifold has a hyperbolic structure.

\paragraph{Topological Invariants.}

Another area of interest, with a number of practical applications
outside mathematics, is the calculation of topological invariants.
Many physical objects can change geometry more easily than
they can change topology.  Examples range from molecules
to alphabetic characters to geological formations.
For these objects, topological invariants 
offer a more meaningful description than geometric measures.

The most useful topological invariants involve {\it homology\/}, which
defines a sequence of groups describing the ``connectedness''  
of a topological space.
For example, the {\it Betti numbers} of an object embedded
in $\R^3$ are respectively the number of connected
components separated by gaps, the number of circles surrounding
tunnels, and the number of shells surrounding voids.
Technically, the Betti numbers are the ranks of the free parts
of the homology groups.
For more abstract topological spaces, not embedded
in $\R^3$, the relevant invariants include {\it torsion
coefficients\/} as well. 

For $2$-manifolds without boundary, the homology can be computed
quite easily by computing Euler characteristics and orientability.
The case of 3-complexes requires more sophistication, but
computational geometers have devised quite efficient algorithms 
for the case of 3-complexes embedded in
$\R^3$~\cite{DelEde-CAGD-95,DeyGuh-JACM-98}. However, these algorithms
use the three-dimensional embedding heavily and it is not yet clear
whether they can be extended to general complexes.  These problems are
not just of mathematical interest: nonmanifold $2$-complexes are used
quite often in modeling shock fronts, crack propagation, or domains made
of two different materials. In dimension beyond 3, there is yet no
algorithm that
would be practical for large complexes.

From a practical point of view it may often be impossible 
to determine the topology of an object completely, and 
estimation of topological invariants may be appropriate.
In materials science, structural properties
of composite materials such as concrete or high-impact plastic
appear to be related to the Betti numbers of randomly selected
cross sections.

Finally, in addition to studying the shape of objects in
space, topological computations may prove useful in studying the
shape of space itself!  Research is currently underway using
astronomical data to investigate the geometry and topology of the universe.  
One approach uses maps of cosmic background radiation
to piece together the global structure of the universe
\cite{CorWee-NAMS-98,Wee-CQG-98}.

\newpage

\section{Techniques}

\hrulefill\par\bigskip\noindent

\noindent

We can already identify a number of techniques
that computational topology could bring to bear on 
the applications described above.
We list them in order from general scientific principles down
to specific algorithmic methods.

\begin{itemize}

\item
{\bf Mathematical Viewpoint.}
Topology separates global shape properties from  local geometric attributes, 
and provides a precise language for discussing these properties. 
Such a language is essential for composing software applications, such
as connecting a mesh generator to a computational fluid dynamics
simulation.
Mathematical abstraction can also unify similar concepts 
from different fields. For example, 
basic questions of robot reachability or molecular docking
become topological questions in the appropriate configuration or 
conformation spaces. 

\item
{\bf Asymptotic Analysis. }
The signature technique of theoretical computer science is asymptotic
worst-case and average-case analysis of algorithms.  This type of
analysis, while sometimes overemphasized as an end in itself, is helpful
in providing a common yardstick to measure progress and encourage future
work.  Although proving upper bounds on algorithm performance is usually a
matter of concrete analysis, topological ideas such as Betti numbers
can be useful in proving lower bounds \cite{Yao-STOC-94}.

\item
{\bf Exact Geometric Computation. }
This technique draws on algebraic number theory
to ensure  the topological correctness of geometric computations.  
In principle, this technique solves most of the numerical
robustness problems in such applications 
as CAD modeling and computational simulation. 

\item
{\bf Differential Methods. }
Many techniques from differential geometry, such 
as Morse theory for studying singularities,
are essential in analyzing surfaces and models
in diverse applications such as medical imaging, crystallography, and
molecular modeling.

\item
{\bf Topological Methods in Discrete Geometry.}
Topological results such as the Borsuk-Ulam theorem
\cite{Bor-FM-33}, that any continuous antipodal function on a sphere
must have a zero, have commonly been used in discrete geometry to prove
the existence of geometric configurations such as ham sandwich cuts and
centerpoints \cite{Bjo-HC-95,Zvi-HDCG-97}.
However, such methods do not generally lead to efficient algorithms
for finding such configurations \cite{Ko-JC-95},
so further research on effective existence proofs may be warranted.

\item
{\bf Multiscale Synthesis and Analysis.}
Multiresolution techniques have already assumed great importance in
the synthesis of computer graphics models and in numerical methods
for physical simulation. 
Multiscale techniques are fast becoming
equally important in visualization and analysis of unstructured ``natural'' data.
One example~\cite{Ler-?,Jon-IM-90} uses
techniques drawn from geometric measure theory and harmonic analysis
for approximating a set with best fit planes at different resolutions.
This approach segments a point set or image into subsets of different
geometric structure; by combining continuous and discrete analysis, it
produces results even for noisy data.

\item
{\bf Normal Surfaces. }
Invented by Kneser~\cite{Kne-JDMV-29}, normal surfaces were the
basis for  Haken's knot algorithm~\cite{Hak-AM-61}, and have been used in
numerous algorithmic and finiteness
results in topology~\cite{math.GT/9712269,math.GT/9811031}.
Instead of representing a curve or surface with an explicit mesh
or parameterization, normal surface theory describes how that
curve or surface intersects a given mesh of the ambient space.
This yields a very efficient representation for densely folded 
curves and surfaces, which has potential in applications
where such curves and surfaces occur.  Moreover, normal surface theory
provides a natural ``addition'' operation for surfaces, which is useful
for their manipulation.

\end{itemize}

\newpage

\section{Recommendations}

\hrulefill\par\bigskip\noindent

\begin{itemize}

\item
{\bf Research Community. }
There is need to build a computational topology
research community including computer scientists,
engineers, and mathematicians.
Such a community could be held together by
organizing workshops and conference
special sessions, and by maintaining Web sites, bibliographies, and
collections of open problems.
Techniques from topology have
already been used in geometric computing and vice versa.
We want to strengthen and formalize this link.

\item
{\bf Online Clearinghouse. }
To encourage a sense of community, we should establish a
clearinghouse of research projects, papers, software, and informal
communications between workers in this area.  The web site already
present at www.computop.org could possibly provide a location for this
collection.

\item
{\bf Research Funding. }
Grant opportunities are needed to encourage further work in
these areas, either as a separate initiative or continued funding from
the relevant areas within NSF.

\item
{\bf Continuation of Workshop. }
It seems premature to establish a journal or annual
conference series in this area, but at the least there should be another
workshop on computational topology.  This year's workshop was an
invitation-only, direction-finding session; what is needed now is a
forum for collecting new work in the area and fostering continued
interdisciplinary collaboration.  Perhaps such an event could be held in
conjunction with the annual ACM Symposium on Computational Geometry, to
be held next year in Hong Kong.
\end{itemize}

\newpage

\section*{Bibliography}
\hrulefill\par\bigskip\noindent

\bibliographystyle{comptop}
\bibliography{report}

\end{document}